\documentclass[11pt]{article}
\usepackage{latexsym}
\usepackage{amssymb}
\usepackage[dvips]{graphicx}
\usepackage{epsfig}
\usepackage{rotating}
\begin{document}
\begin{titlepage}
\begin{flushright}
ICMPA-MPA/2006/20\\
CP3-06-13\\
\end{flushright}

\begin{center}

{\ }\vspace{1cm}

{\Large\bf $(p,q)$-Deformations and $(p,q)$-Vector Coherent States}

\vspace{5pt}

{\Large\bf of the Jaynes-Cummings Model}

\vspace{5pt}

{\Large\bf in the Rotating Wave Approximation}

 \vspace{1.5cm}

Joseph Ben Geloun$^{\dag}$, Jan Govaerts$^{\ddag,\dag,}$\footnote{On sabbatical
leave from the Center for Particle Physics and Phenomenology (CP3),
Institute of Nuclear Physics, Catholic University of Louvain,
2, Chemin du Cyclotron, B-1348 Louvain-la-Neuve, Belgium.}
  and M. Norbert Hounkonnou$^{\dag}$

\vspace{1cm}

$^{\dag}${\em International Chair in Mathematical Physics
and Applications (ICMPA-UNESCO)}\\
{\em 072 B.P. 50 Cotonou, Republic of Benin}\\
{\em E-mail: {\tt jobengeloun@yahoo.fr, norbert$_-$hounkonnou@cipma.net}}

\vspace{1.0cm}

$^{\ddag}${\em Department of Theoretical Physics, School of Physics}\\
{\em The University of New South Wales, Sydney NSW 2052, Australia}\\
{\em E-mail: {\tt Jan.Govaerts@fynu.ucl.ac.be}}

\vspace{1cm}

\today

\begin{abstract}
Classes of $(p,q)$-deformations of the Jaynes-Cummings model in
the rotating wave approximation are considered. Diagonalization of
the Hamiltonian is performed exactly, leading to useful spectral
decompositions of a series of relevant operators. The latter include
ladder operators acting between adjacent energy eigenstates within
two separate infinite discrete towers, except for a singleton state.
These ladder operators allow for the construction of $(p,q)$-deformed
vector coherent states. Using $(p,q)$-arithmetics, explicit and exact
solutions to the associated moment problem are displayed, providing
new classes of coherent states for such models. Finally, in the limit
of decoupled spin sectors, our analysis translates into $(p,q)$-deformations
of the supersymmetric harmonic oscillator, such that the two supersymmetric
sectors get intertwined through the action of the ladder operators as well
as in the associated coherent states.
\end{abstract}

\end{center}

\end{titlepage}

\setcounter{footnote}{0}

\section{Introduction}
\label{Sect1}

In recent years, quantum algebras and groups \cite{maj}
which appear as a generalization of the symmetry concept \cite{wes}
and the  basics of so-called noncommutative theories,
have been the subject of intensive research
interest in both mathematics and physics. The $q$- and
more generally $(p,q)$-deformation of a pre-defined algebraic
structure \cite{ar,far,cha} proves to be a powerful tool widely
used in the representation theory of quantum groups. The field
of ``$q$-mathematics" has a long history \cite{jac,ram} dating
back to over 150 years, and includes several famous names
such as Cauchy, Jacobi and Heine to mention just a few.
Its possible relation to physics has been considerably
reinforced during the last thirty years \cite{ar,hex}.
In particular, great attention has been devoted to
deformations of the bosonic Fock-Heisenberg algebra. The most
commonly studied deformed bosons, with annihilation and creation
operators $a$ and $a^\dagger$, respectively, satisfy the
$q$-commutation relation \cite{ar} (also called quommutation)
\begin{equation}
aa^{\dagger} - q a^{\dagger}a=\mathbb{I},
\label{eq:q-def1}
\end{equation}
or some variant forms of such a relation \cite{far,vin}.
Still more general deformations, which include in specific limits
the above standard $q$-deformed case and which also provides consistent
extensions of the harmonic oscillator algebra, proceed
from the two parameter deformation of the Fock algebra
introduced by Chakrabarty and Jagannathan \cite{cha},
namely the so-called $(p,q)$-oscillator quantum
algebras generated by three operators $a$, $a^{\dagger}$
and $N$ which obey \cite{cha,vin2}
\begin{eqnarray}
[N,a]=-a,\quad
[N,a^{\dag}]=a^{\dag}, \quad
aa^{\dag}-qa^{\dag}a=p^{-N},\quad
aa^{\dag}-p^{-1}a^{\dag}a=q^{N}.
\label{eq:pq-def1}
\end{eqnarray}
Here, $p$ and $q$ are free parameters, which henceforth are chosen to be both
real and such that $p>1$, $0<q<1$ and $pq<1$. Clearly, one recovers the ordinary
Fock algebra of the harmonic oscillator algebra in the double limit $p,q \to 1$,
with then $[a,a^\dagger]=\mathbb{I}$ and $N=a^\dagger a$.
Furthermore, these $q$- and $(p,q)$-deformed algebras have found
a number of relevant applications and provide algebraic
interpretations of various $q$- and $(p,q)$-special functions \cite{vin,vin2,koe}.

The harmonic oscillator algebra is central in the construction
of a number of models in physics, among which the
Jaynes--Cummings model (${\cal JC}m$) plays a significant role.
Indeed ever since Jaynes and Cummings' historical work \cite{jc},
the ${\cal JC}m$ has been at the basis of many investigations.
This system belongs to a class of physically relevant models
widely used in atomic physics and quantum optics.
As far as we know, a great deal of analytically solvable
models of this type have been studied in the rotating
wave approximation (r.w.a.) within the framework of
non-deformed commutative theories (see \cite{jc}--\cite{hus4}
and references therein). The ${\cal JC}m$ has also been considered
in the context of generalized intensity dependent oscillator algebras
including nonlinear dynamical supersymmetry \cite{daou} or
using shape invariance techniques \cite{bal1,bal2}.
Comparatively, much fewer papers have dealt with generalizations of
these models including deformations. Among the latter and
mainly based on the generalized intensity-dependent
coupling of Buck and Sukumar \cite{buk}, one may mention,
on the one hand, the work by Chaichan {\it et al.} \cite{chai},
and on the other hand, that by Chang \cite{chan}, both
dealing with a generalized $q$-deformed intensity-dependent
interaction Hamiltonian of the ${\cal JC}m$ given
by the Holstein-Primakoff $su_q(1,1)$ or $su_q(2)$
quantum algebra realizations of the Hamiltonian field
operators and the related Peremolov, Glauber or Barut-Girardello
group theoretical construction of coherent states.
In the same vein, the paper by Naderi {\it et al.} considers the dynamical
properties of a two-level atom in three variants of the
two-photon $q$-deformed ${\cal JC}m$ \cite{nad}. In this latter work,
the authors focused their attention onto
the time evolution of atomic properties including population inversion
and quantum fluctuations of the atomic dipole variables.
However, it is not clear to us how the main issues related to the
moment problem as well as the mathematical foundation of the coherent
and squeezed states which they use and on which a great part of their
analysis rests in a crucial way, are solved.

In a recent publication \cite{hus1}, Hussin and Nieto have performed
an interesting systematic search of different types of ladder operators
for the ${\cal JC}m$ model in the r.w.a. and constructed
associated coherent states. In the present work, and in
line with that investigation, we provide a generalization
of that analysis to $(p,q)$-deformations of the same model.

The outline of the paper is the following. In Section \ref{Sect2},
we briefly recall the main results relevant to
the ${\cal JC}m$ in the r.w.a. in the non-deformed situation \cite{hus1}.
Section \ref{Sect3} then introduces $(p,q)$-deformations of the same
model. By providing an explicit diagonalization of the $(p,q)$-deformed
Hamiltonian, the spectrum and its eigenstates are exactly identified.
As in the non-deformed case \cite{hus1}, except for a singleton state,
all other energy eigenstates are organized into two separate discrete towers,
for which ladder operators transforming states into one another within
each tower separately may be introduced. Using properties of these
ladder operators, in Section \ref{Sect4} we introduce general classes of $(p,q)$-deformed
vector coherent states. The freedom afforded in their construction
is fixed from two alternative points of view, discussed in Section \ref{Sect5},
which in the ordinary case of the non-deformed Fock algebra coincide. However at all stages of our
discussion, the double limit $p,q\to 1$ reproduces the corresponding results of
\cite{hus1}. Section \ref{Sect5} also briefly considers the situation in the uncoupled
limit of the ${\cal JC}m$, while Section \ref{Sect6} presents some concluding remarks.
An Appendix collects useful facts in connection with properties of
$(p,q)$-deformed algebras and related functions.

\section{The Ordinary ${\cal JC}m$ in the Rotating Wave Approximation}
\label{Sect2}

The ${\cal JC}m$ describes the interaction between one mode of
the quantized electromagnetic field and a two-level model of an atomic system
\cite{jc,hus1}--\cite{hus3}. It has proved to be a theoretical
laboratory of great relevance to many topics in atomic physics
and quantum optics, as well as in the  study of ion traps,
cavity QED theory and quantum information processing \cite{mew,hus1}.
Furthermore, the spin-orbit interaction term which
appears in the ${\cal JC}m$ is essentially the
so-called Dresselhaus spin-orbit term \cite{dr}. The model is thus
also widely used in condensed matter physics for its relevance
in spintronics \cite{js} which exploits the electron spin rather than
its charge to develop a new generation of electronic devices \cite{qui1,qui2}.
The solution of the complete ${\cal JC}m$ is not
yet known in a closed form \cite{hus1}. However,
in the r.w.a., although the Hamiltonian remains
nonlinear, the model becomes exactly solvable in closed form
with explicit expressions for its eigenenergy states.
In this Section, we briefly recall, in a streamlined presentation,
the main results in the non-deformed case (see \cite{hus1,hus2} and references therein)
of relevance to our analysis of $(p,q)$-deformations hereafter.

In the r.w.a., the reduced dimensionless ${\cal JC}m$ Hamiltonian reads \cite{hus2}
\begin{eqnarray}
\label{h}
{\cal H}^{\rm red}=\frac{1}{\hbar \omega_0}{\cal H}=
\left(1 + \epsilon\right)\left(a^\dagger a + \frac{1}{2}\right) +
\frac{1}{2}\sigma_{3}+\lambda\left(a^\dagger\sigma_{-} +a \sigma_{+}\right),
\end{eqnarray}
where $a$ and $a^\dagger$ are the usual photon annihilation and creation
operators, respectively, obeying the ordinary Fock algebra,
and $(\sigma_1,\sigma_2,\sigma_3)$ are the
Pauli matrices with $\sigma_\pm= \sigma_1 \pm i\sigma_2$.
The r.w.a. is related to the detuning parameter $\epsilon$ which is
such that $|\epsilon|\ll 1$, with $\omega_0$ being the fixed atomic
frequency and $\omega=\omega_0(1+\epsilon)$ the actual field mode frequency.
The r.w.a. is reliable provided $|\omega - \omega_0|\ll \omega,\omega_0$.
Finally, $\lambda$ is the reduced spin-orbit coupling modelling the interaction
strength between the radiation field and the atom.

\clearpage

The Hilbert space ${\cal V}$ of the system is the tensor product of the Fock space
representation of the Fock algebra $(a,a^\dagger)$ and the 2-dimensional
representation of the SU(2) algebra associated to the Pauli matrices.
A basis of the former is provided by the number operator, $N=a^\dagger a$,
orthonormalized eigenstates
$|n\rangle=(1/\sqrt{n!})(a^\dagger)^n|0\rangle$ ($n=0,1,2,\cdots$),
with $a|n\rangle=\sqrt{n}|n-1\rangle$, $a^\dagger|n\rangle=\sqrt{n+1}|n+1\rangle$
and $N|n\rangle=n|n\rangle$, while a basis of the latter spin sector is the orthonormalized set
$\{|+\rangle,|-\rangle\}$ such that $\sigma_3|\pm\rangle=\pm|\pm\rangle$.
The tensor product space is thus spanned by the states $|n,\pm\rangle=|n\rangle\otimes|\pm\rangle$.

The diagonalization of the Hamiltonian (\ref{h}) is readily achieved.
The orthonormalized energy eigenspectrum consists of a ``singleton" state $|E_*\rangle$,
\begin{equation}
{\cal H}^{\rm red}|E_*\rangle= E_*|E_*\rangle,
\end{equation}
with
\begin{equation}
E_*=\frac{1}{2}\epsilon,\qquad
|E_*\rangle=|0,-\rangle ,
\end{equation}
and two infinite discrete towers of states $|E_n^\pm\rangle$ such that
${\cal H}^{\rm red}|E^\pm_n\rangle=E^\pm_n|E^\pm_n\rangle$ for all
$n=0,1,2,\cdots$, expressed as \cite{hus1}
\begin{eqnarray}
\label{ei}
|E^+_{n}\rangle&=&\sin\vartheta(n)\, |n,+\rangle + \cos\vartheta(n)\, |n+1,-\rangle ,\\
|E^-_{n}\rangle&=&\cos\vartheta(n)\, |n,+\rangle - \sin\vartheta(n)\, |n+1,-\rangle,
\end{eqnarray}
where, given $Q(n+1)=\sqrt{\epsilon^{2}/4 + \lambda^2(n+1)}$,
the mixing angle $\vartheta(n)$ is such that
\begin{eqnarray}
\label{de}
\sin\vartheta(n)= {\rm sign}(\lambda)
\sqrt{\frac{Q(n+1) - \epsilon/2}{2Q(n+1)}},\qquad
\cos\vartheta(n)= \sqrt{\frac{Q(n+1) + \epsilon/2}{2Q(n+1)}},
\end{eqnarray}
while the energy eigenvalues are
\begin{eqnarray}
\label{ev}
E^{\pm}_{n} = (1+ \epsilon)(n+1) \pm Q(n+1).
\end{eqnarray}
Consequently, one has the spectral decomposition of the reduced Hamiltonian (\ref{h}),
\begin{equation}
{\cal H}^{\rm red}=|E_*\rangle\,E_*\,\langle E_*|\ +\
\sum_{n=0,\pm}^\infty\,|E^\pm_n\rangle\,E^\pm_n\,\langle E^\pm_n| .
\end{equation}

It proves useful to introduce the following notations. Let ${\cal V}_0$ be the (complex)
one-dimensional subspace of the Hilbert space ${\cal V}$ spanned by the state $|0,-\rangle=|E_*\rangle$,
and $\overline{\cal V}$ be its complement in the Hilbert space ${\cal V}$, spanned by
$\{|E^\pm_n\rangle, n\in\mathbb{N}\}$. We thus have ${\cal V}={\cal V}_0\oplus\overline{\cal V}$.

Furthermore let us introduce \cite{hus1} operators ${\cal U}$ and
${\cal U}^\dagger$ defined through their action on the above two sets of basis vectors,
for all $n\in\mathbb{N}$,
\begin{equation}
{\cal U}|n,\pm\rangle = |E^\pm_n\rangle ;\qquad
{\cal U}^\dagger|E_*\rangle=0,\quad
{\cal U}^\dagger|E^\pm_n\rangle=|n,\pm\rangle,
\end{equation}
namely
\begin{equation}
{\cal U}=\sum_{n=0,\pm}^\infty |E^\pm_n\rangle\langle n,\pm|,\qquad
{\cal U}^\dagger=\sum_{n=0,\pm}^\infty |n,\pm\rangle\langle E^\pm_n|.
\end{equation}
Clearly we have
\begin{equation}
{\cal U}\,{\cal V}=\overline{\cal V};\qquad
{\cal U}^\dagger\,{\cal V}={\cal V},\quad
{\cal U}^\dagger\,\overline{\cal V}={\cal V}.
\end{equation}
Note that even though neither ${\cal U}$ nor ${\cal U}^\dagger$ is unitary on the full Hilbert space
${\cal V}$, they are the adjoint of one another, hence the notation.

It is of interest to apply these operators onto the quantum Hamiltonian (\ref{h}). One obtains
\begin{equation}
\mathbb{H}^{\rm red}={\cal U}^\dagger\,{\cal H}^{\rm red}\,{\cal U}=
\sum_{n=0,\pm}^\infty\,|n,\pm\rangle\,E^\pm_n\,\langle n,\pm|,
\end{equation}
and conversely,
\begin{equation}
{\cal U}\,\mathbb{H}^{\rm red}\,{\cal U}^\dagger=
\sum_{n=0,\pm}^\infty |E^\pm_n\rangle\,E^\pm_n\,\langle E^\pm_n|=
{\cal H}^{\rm red}\,-\,|E_*\rangle\,E_*\,\langle E_*|.
\end{equation}

The energy eigenstates spanning $\overline{\cal V}$ may be organized
into two subspaces referred to as ``towers", namely
$\left\{|E^+_n\rangle, n\in\mathbb{N}\right\}$ and
$\left\{|E^-_n\rangle, n\in\mathbb{N}\right\}$.
The states in the tower  $\left\{|E^+_n\rangle, n\in\mathbb{N}\right\}$
are associated to strictly increasing eigenvalues
so that they constitute a nondegenerate set of eigenstates.
The second group does not necessarily possess the same feature
depending on the values for the parameters $\lambda$ and $\epsilon$.
It is possible \cite{hus3} to identify a range of values for these
parameters such that $\left\{|E^-_n\rangle, n\in\mathbb{N}\right\}$
only contains nondegenerate states of strictly increasing eigenvalues with $n$.
Some of the considerations discussed hereafter may require a nondegenerate
spectrum, which may always be achieved by properly ``detuning" the parameters
$\lambda$ and $\epsilon$ away from a degenerate case, but not necessarily a
strictly increasing spectrum in the label $n\in\mathbb{N}$. Whatever the
case may be though, bounded from below spectra such that $E^\pm_n>E^\pm_0$ for
$n=1,2,\cdots$ are always assumed implicitly.

It is possible to consider ladder operators acting between successive
energy eigenstates within each of the above two towers, irrespective of
whether the spectral values are strictly increasing or not\footnote{We
differ on this point with \cite{hus1}, where strictly increasing energy
spectra in each tower are required.}. Namely, let us first consider
operators $\mathbb{M}^-$ and $\mathbb{M}^+$ given as
\begin{equation}
\mathbb{M}^-=\sum_{n=0,\pm}^\infty|n-1,\pm\rangle\,K_\pm(n)\,\langle n,\pm|;\qquad
\mathbb{M}^+=\sum_{n=0,\pm}^\infty|n+1,\pm\rangle\,K^*_\pm(n+1)\,\langle n,\pm|,
\end{equation}
where $K_\pm(n)$ are, at this stage, arbitrary complex coefficients such that $K_\pm(0)=0$.
Then, introduce the ladder operators
\begin{equation}
{\cal M}^-={\cal U}\,\mathbb{M}^-\,{\cal U}^\dagger=
\sum_{n=0,\pm}^{\infty}|E^\pm_{n-1}\rangle\,K_\pm(n)\,\langle E^\pm_n|;\quad
{\cal M}^+={\cal U}\,\mathbb{M}^+\,{\cal U}^\dagger=
\sum_{n=0,\pm}^{\infty}|E^\pm_{n+1}\rangle\,K^*_\pm(n+1)\,\langle E^\pm_n|,
\end{equation}
which are thus such that, for all $n=0,1,2,\cdots$,
\begin{equation}
{\cal M}^-|E_*\rangle=0,\quad
{\cal M}^-|E^\pm_n\rangle=K_\pm(n)|E^\pm_{n-1}\rangle;\quad
{\cal M}^+|E_*\rangle=0,\quad
{\cal M}^+|E^\pm_n\rangle=K^*_\pm(n+1)|E^\pm_{n+1}\rangle .
\label{eq:ladder1}
\end{equation}
Note that ${\cal M}^-$ and ${\cal M}^+$ are adjoint of one another
but in effect only act on the subspace $\overline{\cal V}$.

General vector coherent states (VCS) may then be introduced \cite{al}--\cite{jp2}
on the space $\overline{\cal V}$ as eigenstates of the lowering operator ${\cal M}^-$ with
as eigenvalue an arbitrary complex number $z\in\mathbb{C}$. Furthermore, these VCS are
also parametrized by two real quantities $\tau_\pm$ which account for their stability
under time evolution generated by the operator $\exp\left\{-i\omega_0 t\,{\cal H}^{\rm red}\right\}$,
as well as the two spherical coordinates $(\theta,\phi)\in[0,\pi]\times[0,2\pi[$
parametrizing a unit vector in the 2-sphere $S_2$ (hence the name of ``vector" coherent states).
Explicitly, one has \cite{hus1}
\begin{eqnarray}
\label{cohp}
|z;\tau_\pm;\theta,\phi\rangle&=&
\ \ \ N^+(|z|)\cos\theta \sum_{n=0}^\infty
\frac{z^n}{K_+(n)!} e^{-i \omega_0\tau_+ E^+_n}\,|E^+_n\rangle \cr
& & + \ N^-(|z|)\, e^{i\phi}\sin\theta \sum_{n=0}^\infty
\frac{z^n}{K_-(n)!} e^{-i \omega_0\tau_- E^-_n}\,|E^-_n\rangle ,
\end{eqnarray}
where $K_\pm(n)!=\prod_{k=1}^n K_\pm(k)$ (with, by convention, $K_\pm(0)!=1$),
while the normalization factors are defined as
\begin{equation}
N^{\pm}(|z|)= \left[\sum_{n=0}^\infty
\frac{|z|^{2n}}{|K_\pm(n)!|^2}\right]^{-1/2}
\end{equation}
in order that the VCS be of unit norm. The smallest value, $R$, of the two convergence radii of
these two series in $|z|$ also defines the disk $D_R$ in $z\in\mathbb{C}$ for which
these VCS are well defined. These states are clearly such that
\begin{equation}
{\cal M}^-|z;\tau_\pm;\theta,\phi\rangle = z\, |z;\tau_\pm;\theta,\phi\rangle ,\quad
e^{-i\omega_0 t\,{\cal H}^{\rm red}}\,|z;\tau_\pm;\theta,\phi\rangle=|z;t+\tau_\pm;\theta,\phi\rangle.
\end{equation}

Further restrictions are necessary to finally specify in a unique fashion the factors $K_\pm(n)$,
and then solve the moment problem implied by the requirement of overcompleteness over
$\overline{\cal V}$ for the VCS (\ref{cohp}) given a choice of a SU(2) matrix-valued integration
measure over $\mathbb{C}\times S_2$ \cite{kl}-\cite{jp2}. Different choices are
available \cite{hus1}, each leading to a different set of VCS. Furthermore,
taking the limit case $\lambda  \to 0$ or the zero-detuning limit (resonance case)
$\epsilon \to 0$, different models arise with their associated VCS.

For the sake of illustration, let us consider one such choice explicitly \cite{hus1}.
The factors $K_\pm(n)$ may be restricted for example by requiring that the
ladder operators ${\cal M}^-$ and ${\cal M}^+$ obey the usual Fock algebra
of annihilation and creation operators on the space $\overline{\cal V}$,
\begin{equation}
\left[{\cal M}^-,{\cal M}^+\right]=
{\cal M}^-\,{\cal M}^+\,-\,{\cal M}^+\,{\cal M}^-\,=\,\mathbb{I}_{\overline{\cal V}}
=\sum_{n=0,\pm}^\infty\,|E^\pm_n\rangle\,\langle E^\pm_n|.
\end{equation}
{}From the expressions in (\ref{eq:ladder1}) and the initial conditions $K_\pm(0)=0$,
it follows that the quantities $K_\pm(n)$ are
now determined up to arbitrary phase factors $\varphi_\pm(n)$ as
\begin{equation}
K_\pm(n)=e^{i\varphi_\pm(n)}\,\sqrt{n},\qquad n=0,1,2,\cdots .
\end{equation}
Consequently, one has $N^\pm(|z|)=e^{-|z|^2/2}$, which is well-defined
for all $z\in\mathbb{C}$. Hence so are then all the VCS $|z;\tau_\pm;\theta,\phi\rangle$.

\section{The $(p,q)$-Deformed ${\cal JC}m$ in the Rotating Wave Approximation}
\label{Sect3}

Let us now introduce a $(p,q)$-deformation of the ${\cal JC}m$ Hamiltonian (\ref{h}),
namely $(p,q)$-${\cal JC}m$ models. The eigenstates and spectrum are first identified,
before considering the construction of ladder operators following the same rationale
as in Section \ref{Sect2}. A study of the associated VCS and examples of exactly
solvable reduced models is differed to Section \ref{Sect4}.

\subsection{Energy spectrum and eigenstates}
\label{Subsect3.1}

Given the $(p,q)$-deformation (\ref{eq:pq-def1}) of
the ordinary Fock algebra (see the Appendix for further details
and identities pertaining to such deformations), we now consider
$(p,q)$-deformations of the Hamiltonian (\ref{h}) of the form\footnote{Make no
mistake that henceforth, all quantities correspond to the $(p,q)$-deformed analysis
even though the notations used coincide with those of Section \ref{Sect2}
and do not make explicit the fact that all expressions correspond now to the
deformed case. When wanting to make the difference explicit, notations such as
for instance $[N]\equiv [N]_{(p,q)}=(p^{-N}-q^N)/(p^{-1}-q)$
and $[n]\equiv [n]_{(p,q)}=(p^{-n}-q^n)/(p^{-1}-q)$ are used.}
\begin{eqnarray}
{\cal H}^{red}= \left(1+\epsilon\right)\left\{h(p,q)[N] + \frac{1}{2} \right\}
 + \frac{1}{2}\sigma_3 + \lambda\left(a^\dagger\sigma_- + a \sigma_+\right),
\label{eq:hq}
\end{eqnarray}
where $[N]=(p^{-N}-q^N)/(p^{-1}-q)$, and $h(p,q)$ is some arbitrary positive function
of the real parameters $p>1$ and $0<q<1$ (with $pq<1$) such that $\lim_{p,q\to 1}h(p,q)=1$
in order to recover (\ref{h}) in the non-deformed case.

The Hilbert space ${\cal V}$ of quantum states of the model is again the tensor product of
the $(p,q)$-deformed Fock space spanned by the states\footnote{Once again, the states
$|n\rangle=|n\rangle_{(p,q)}$ are not to be confused with the number operator eigenstates
of the ordinary Fock algebra as in Section \ref{Sect2}, in spite of an identical notation.}
$|n\rangle$ ($n\in\mathbb{N}$) such as $a|n\rangle=\sqrt{[n]}|n-1\rangle$ and
$a^\dagger|n\rangle=\sqrt{[n+1]}|n+1\rangle$ (see the Appendix), with the 2-dimensional
representation of the SU(2) algebra associated to the Pauli matrices $\sigma_i$ ($i=1,2,3$).
Hence the diagonalization of (\ref{eq:hq}) is readily achieved in the same way
as in the non-deformed case, on the basis
$|n,\pm\rangle=|n\rangle\otimes|\pm\rangle$ of ${\cal V}$.

For any $n\in\mathbb{N}$, let us introduce the following quantities,
\begin{equation}
{\cal E}([n+1])=\left(1+\epsilon\right)h(p,q)\Big([n+1]-[n]\Big)-1,\quad
Q([n+1])=\sqrt{\frac{1}{4}{\cal E}^2([n+1])\,+\,\lambda^2\,[n+1]},
\label{eq:EQ}
\end{equation}
as well as the mixing angles $\vartheta([n])$ defined by
\begin{equation}
\sin\vartheta([n])={\rm sign}(\lambda)
\sqrt{\frac{Q([n+1])-{\cal E}([n+1])/2}{2Q([n+1])}},\quad
\cos\vartheta([n])=
\sqrt{\frac{Q([n+1])+{\cal E}([n+1])/2}{2Q([n+1])}}.
\label{eq:mixang}
\end{equation}
The energy eigenspectrum of (\ref{eq:hq}) is then obtained as follows.
First, there exists a singleton state $|E_*\rangle=|0,-\rangle$ such that
\begin{equation}
{\cal H}^{\rm red}\,|E_*\rangle=E_*\,|E_*\rangle,\qquad
E_*=\frac{1}{2}\epsilon,
\end{equation}
with an eigenvalue which is thus independent of the deformation
parameters $p$ and $q$. Next, one also finds two infinite discrete
towers of states for all $n\in\mathbb{N}$ such that
\begin{eqnarray}
|E^+_n\rangle &=& \sin\vartheta([n])\,|n,+\rangle\,+\,\cos\vartheta([n])\,|n+1,-\rangle ,\\
|E^-_n\rangle &=& \cos\vartheta([n])\,|n,+\rangle\,-\,\sin\vartheta([n])\,|n+1,-\rangle ,
\end{eqnarray}
with
\begin{equation}
{\cal H}^{\rm red}\,|E^\pm_n\rangle= E^\pm_n\,|E^\pm_n\rangle ,\quad
E^\pm_n=\frac{1}{2}\left(1+\epsilon\right)
\Big\{h(p,q)\Big([n+1]+[n]\Big)+1\Big\}\,\pm\,Q([n+1]).
\label{eq:pqev}
\end{equation}
Note that the energy spectrum of these states is deformed by the parameters
$p$ and $q$ as compared to the ordinary case. In particular, the Zeeman spin splitting
$\Delta E_n=E^+_n-E^-_n=2Q([n+1])$, proportional to the Rabi frequency,  is function of the values for
$p$ and $q$. In terms of these results, the reduced Hamiltonian (\ref{eq:hq}) possesses
the spectral resolution
\begin{equation}
{\cal H}^{\rm red}=|E_*\rangle\,E_*\,\langle E_*|\ +\,
\sum_{n=0,\pm}^\infty\,|E^\pm_n\rangle\,E^\pm_n\,\langle E^\pm_n| .
\end{equation}

Let us again introduce the following notations and operators. Let ${\cal V}_0$
denote the subspace of the Hilbert space ${\cal V}$ spanned by the singleton state
$|E_*\rangle=|0,-\rangle$, and $\overline{\cal V}$ its complement in ${\cal V}$,
namely the subspace spanned by $\{|E^\pm_n\rangle,n\in\mathbb{N}\}$, with of course
${\cal V}={\cal V}_0\oplus\overline{\cal V}$. Acting on these spaces, let us
consider the operators
\begin{equation}
{\cal U}=\sum_{n=0,\pm}^\infty\,|E^\pm_n\rangle\,\langle n,\pm|;\qquad
{\cal U}^\dagger=\sum_{n=0,\pm}^\infty\,|n,\pm\rangle\,\langle E^\pm_n|,
\end{equation}
such that, for all $n=0,1,2,\cdots$,
\begin{equation}
{\cal U}|n,\pm\rangle=|E^\pm_n\rangle ;\quad
{\cal U}^\dagger|E_*\rangle=0,\quad
{\cal U}^\dagger|E^\pm_n\rangle=|n,\pm\rangle,
\end{equation}
and thus
\begin{equation}
{\cal U}\,{\cal V}=\overline{\cal V};\quad
{\cal U}^\dagger\,{\cal V}={\cal V},\quad
{\cal U}^\dagger\,\overline{\cal V}={\cal V}.
\end{equation}
Hence once again the operators ${\cal U}$ and ${\cal U}^\dagger$, even though
non unitary on ${\cal V}$, are adjoint of one another. More specifically, one has
\begin{equation}
{\cal U}^\dagger\,{\cal U}=\sum_{n=0,\pm}^\infty\,|n,\pm\rangle\,\langle n,\pm|=\mathbb{I}_{\cal V},\qquad
{\cal U}\,{\cal U}^\dagger=\sum_{n=0,\pm}^\infty\,|E^\pm_n\rangle\,\langle E^\pm_n|=\mathbb{I}_{\overline{\cal V}}.
\end{equation}

Applying these operators to the reduced Hamiltonian, one finds
\begin{equation}
\mathbb{H}^{\rm red}={\cal U}^\dagger\,{\cal H}^{\rm red}\,{\cal U}=
\sum_{n=0,\pm}^\infty\,|n,\pm\rangle\,E^\pm_n\,\langle n,\pm|,
\label{eq:hpq}
\end{equation}
and conversely,
\begin{equation}
{\cal U}\,\mathbb{H}^{\rm red}\,{\cal U}^\dagger=
\sum_{n=0,\pm}^\infty\,|E^\pm_n\rangle\,E^\pm_n\,\langle E^\pm_n|=
{\cal H}^{\rm red}\,-\,|E_*\rangle\,E_*\,\langle E_*| .
\end{equation}

Some remarks on the spectrum are in order.
First, as in the ordinary ${\cal JC}m$, except for the singleton state
$|E_*\rangle=|0,-\rangle$, the spectrum is the direct sum of two towers
of states $\{|E^\pm_n\rangle, n\in\mathbb{N}\}$. However, in contradistinction
to the non-deformed case or even the $q$-deformation with $p=1$,
the $(p,q)$-basic numbers $[n]=[n]_{(p,q)}$ are not strictly increasing
as a function of $n \in {\mathbb{N}}$ when $p>1$, $0<q<1$ and $pq<1$.
There always exists a finite positive value $n_0\in\mathbb{N}$ such that
$[n]$ decreases once $n>n_0$. Hence, depending on
the values for the parameters $\lambda$ and $\epsilon$ as well as the
positive function $h(p,q)$, parts of the spectrum $E^\pm_n$ may turn negative
or present some degeneracies (as in \cite{hus3}). Without exploring this issue
any further in the present work, henceforth we shall assume that parameter
values are such that no degeneracies occur and that the spectrum $E^-_n$
remains bounded from below ($E^+_n$ is obviously positive).
The definition of the ladder operators to be considered next does not
require a strictly increasing spectrum, while it is only for one of
possible choices leading to vector coherent states to be discussed hereafter
that the condition of non degeneracy in $E^\pm_n>E^\pm_0$, for $n\ge 1$, becomes relevant.
Since it has been shown \cite{hus3} that such conditions may be met
in the non-deformed case for appropriate ranges of values for the available
parameters, through an argument of continuity in the deformation parameters
$p$ and $q$, similar ranges ought to exist also for the $(p,q)$-deformed
realizations of the ${\cal JC}m$ model.

Another feature of potential interest related to these facts, and which
will also not be pursued here, is the possibility that through
the $(p,q)$-deformation of the ${\cal JC}m$, the levels
$E^+_n$ and $E^-_{n+1}$ cross one another. Such a property may lead
to effects similar to the phenomenon of resonant spin-Hall conductance
at the Fermi level recently observed in spintronics \cite{qui1,qui2}.
Note that this $(p,q)$-dependent crossing phenomenon is
expected since the Zeeman splitting $\Delta E_n$ is also modified as a function
of $p$ and $q$. This remark is also in line with the recent
suggestion \cite{Scholtz1,Scholtz2,JBG1} that $(p,q)$-deformed or space noncommutative
realizations of exactly solvable systems may provide useful model approximations
to more realistic complex interacting dynamics of collective phenomena.

\subsection{Ladder operators}
\label{Subsect3.2}

In order to construct ladder operators mapping each of the
successive states $|E^\pm_n\rangle$ into one another separately
within each of the towers, let us first introduce the following
operators acting on ${\cal V}$,
\begin{equation}
\mathbb{A}^-=\sum_{n=0,\pm}^\infty\,|n-1,\pm\rangle\,K_\pm([n])\,
\langle n,\pm|;\quad
\mathbb{A}^+=\sum_{n=0,\pm}^\infty\,|n+1,\pm\rangle\,K^*_\pm([n+1])\,
\langle n,\pm|,
\end{equation}
where $K_\pm([n])$ are arbitrary complex quantities such that
$K_\pm([0])=K_\pm(0)=0$. Note that $\mathbb{A}^-$ and
$\mathbb{A}^+$ are adjoint of one another on ${\cal V}$.

Then the relevant ladder operators are obtained as
\begin{equation}
{\cal A}^-={\cal U}\,\mathbb{A}^-\,{\cal U}^\dagger=
\sum_{n=0,\pm}^\infty\,|E^\pm_{n-1}\rangle\,K_\pm([n])\,\langle E^\pm_n|;\quad
{\cal A}^+={\cal U}\,\mathbb{A}^+\,{\cal U}^\dagger=
\sum_{n=0,\pm}^\infty\,|E^\pm_{n+1}\rangle\,K^*_\pm([n+1])\,\langle E^\pm_n|.
\end{equation}
Consequently, we have indeed, for all $n\in\mathbb{N}$,
\begin{equation}
{\cal A}^-|E_*\rangle=0,\quad
{\cal A}^-|E^\pm_n\rangle=K_\pm([n])\,|E^\pm_{n-1}\rangle;\qquad
{\cal A}^+|E_*\rangle=0,\quad
{\cal A}^+|E^\pm_n\rangle=K^*_\pm([n+1])\,|E^\pm_{n+1}\rangle.
\end{equation}
Note that ${\cal A}^-$ and ${\cal A}^+$ are adjoint of one another,
but that in effect they act only on the subspace $\overline{\cal V}$.

It is of course possible to express these ladder operators in the
$|n,\pm\rangle$ basis. In the case of the lowering operator, one finds
\begin{equation}
\begin{array}{rclcl}
{\cal A}^-&=& \ \ \ \ \sum_{n=0}^\infty\,|n,+\rangle\,{\cal A}^-_{++}(n)\,\langle n+1,+| \ &+&\
\sum_{n=0}^\infty\,|n,+\rangle\,{\cal A}^-_{+-}(n)\,\langle n+2,-| \\
 && && \\
 && +\ \sum_{n=0}^\infty\,|n,-\rangle\,{\cal A}^-_{-+}(n)\,\langle n,+| \ &+&\
\sum_{n=0}^\infty\,|n,-\rangle\,{\cal A}^-_{--}(n)\,\langle n+1,-|
\end{array}
\end{equation}
where
\begin{eqnarray}
{\cal A}^-_{++}(n)&=&\sin\vartheta([n])\,\sin\vartheta([n+1])\,K_+([n+1])\,+\,
\cos\vartheta([n])\,\cos\vartheta([n+1])\,K_-([n+1]),\cr
 & & \cr
{\cal A}^-_{+-}(n)&=&\sin\vartheta([n])\,\cos\vartheta([n+1])\,K_+([n+1])\,-\,
\cos\vartheta([n])\,\sin\vartheta([n+1])\,K_-([n+1]),\cr
 & & \cr
{\cal A}^-_{-+}(n)&=&\cos\vartheta([n-1])\,\sin\vartheta([n])\,K_+([n])\,-\,
\sin\vartheta([n-1])\,\cos\vartheta([n])\,K_-([n]),\cr
 & & \cr
{\cal A}^-_{--}(n)&=&\cos\vartheta([n-1])\,\cos\vartheta([n])\,K_+([n])\,+\,
\sin\vartheta([n-1])\,\sin\vartheta([n])\,K_-([n]).
\end{eqnarray}
Likewise for the raising operator,
\begin{equation}
\begin{array}{rclcl}
{\cal A}^+&=& \ \ \ \ \sum_{n=0}^\infty\,|n+1,+\rangle\,\left({\cal A}^{-}_{++}(n)\right)^*\,\langle n,+| \ &+&\
\sum_{n=0}^\infty\,|n,+\rangle\,\left({\cal A}^{-}_{-+}(n)\right)^*\,\langle n,-| \\
 && && \\
 && +\ \sum_{n=0}^\infty\,|n+2,-\rangle\,\left({\cal A}^{-}_{+-}(n)\right)^*\,\langle n,+| \ &+&\
\sum_{n=0}^\infty\,|n+1,-\rangle\,\left({\cal A}^{-}_{--}(n)\right)^*\,\langle n,-|.
\end{array}
\end{equation}
Note that we have ${\cal A}^-_{-+}(0)=0={\cal A}^-_{--}(0)$, since $K_\pm([0])=0$.

The quantities $K_\pm([n])$ parametrize the freedom available in the choice
of such ladder operators. Further restrictions arise when considering first
the possible existence of vector coherent states meeting a series of general
conditions charateristic of such states \cite{kl}-\cite{jp2}, starting with
one involving the lowering operator ${\cal A}^-$ itself.

\section{$(p,q)$-Vector Coherent States for the $(p,q)$-${\cal JC}m$}
\label{Sect4}

By considering the action of the lowering operator ${\cal A}^-$,
we are able to construct an overcomplete set of vectors in $\overline{\cal V}$,
so-called vector coherent states \cite{kl}-\cite{jp2}
for the $(p,q)$-${\cal JC}m$. Since these states are associated
to unit vectors in the 2-sphere $S_2$ \cite{al},
they are referred to as $(p,q)$-vector coherent states ($(p,q)$-VCS).
As in Section \ref{Sect2}, these $(p,q)$-VCS are parametrized by
a complex variable $z\in\mathbb{C}$, two real parameters $\tau_\pm$
to track a stable time evolution of the $(p,q)$-VCS, and finally
the spherical angle coordinates $(\theta,\phi)$ on $S_2$,
$|z;\tau_\pm;\theta,\phi\rangle$.
In the double limit that $p,q\to 1$, these $(p,q)$-VCS reduce to
those of \cite{hus1} discussed in Section \ref{Sect2}.
The dependence of the $(p,q)$-VCS on all these quantities is
introduced as follows, according to the discussion in \cite{kl}.

\subsection{Identifying $(p,q)$-VCS}
\label{Subsect4.1}

As a slight extension of the analysis so far,
given two real parameters $\mu$ and $\nu$, let us consider the operator
\begin{equation}
\mathbb{Q}_{\cal V}=|E_*\rangle\,\langle E_*|\ +\
\sum_{n=0,\pm}^\infty\,|E^\pm_n\rangle\,
\left(\frac{q^{\mu}}{p^{\nu}}\right)^{n}
\,\langle E^\pm_n|.
\end{equation}
Hence, the energy eigenstates of the $(p,q)$-${\cal JC}m$ are
also eigenstates of this operator $\mathbb{Q}_{\cal V}$,
with eigenvalues given through the above spectral decomposition.

We are now in a position to successively identify the dependence
of the $(p,q)$-VCS to be constructed on each of the parameters
of which they are functions, first $z$, then $\tau_\pm$,
and finally, $\theta$ and $\phi$. Having defined both the operators
${\cal A}^-$ and $\mathbb{Q}_{\cal V}$, let us consider the following
eigenvalue problem in $z$ for the $(p,q)$-VCS,
\begin{eqnarray}
\label{coh}
{\cal A}^-|z;\tau_\pm;\theta,\phi\rangle =
 z\,\mathbb{Q}_{\cal V}\,|z;\tau_\pm;\theta,\phi\rangle
\end{eqnarray}
which generalizes to a two-level system the definition of coherent
states as advocated in \cite{kl}-\cite{jp2}. The particular case $\mu=0=\nu$
yields also a consistent definition of $(p,q)$-VCS viewed as
the limit $\mu,\nu \to 0$ of the present definition (note that their domain
of definition in $z$, required for the convergence of the infinite series to be
considered hereafter, may have to be adapted accordingly).

By expanding the $(p,q)$-VCS in the Hamiltonian eigenstate basis as
\begin{eqnarray}
\label{ser}
|z;\tau_\pm;\theta,\phi\rangle=
C_*(z)|E_*\rangle + \sum_{n=0,\pm}^\infty\,C^\pm_n(z)|E^\pm_n\rangle,
\end{eqnarray}
where $C_*(z)$ and $C^\pm_n(z)$ are complex continuous functions of $z$
to be specified presently, the condition (\ref{coh}) then requires,
for all $n\in\mathbb{N}$,
\begin{eqnarray}
\label{cc}
C_*(z)=0,\qquad
C^\pm_{n+1}(z)K_\pm([n+1]) = z\,\frac{q^{\mu n}}{p^{\nu n}}\,C^\pm_n(z),
\end{eqnarray}
of which the solution is
\begin{eqnarray}
C^\pm_n(z)= \left(\frac{q^{\mu}}{p^{\nu}}\right)^{n(n-1)/2}\,
\frac{z^n}{K_\pm([n])!}\,C^\pm_0(z),
\end{eqnarray}
where $C^\pm_0(z)$ are arbitrary complex functions of $z$, while we
defined $K_\pm([n])!=\prod_{k=1}^n K_\pm([k])$ with, by convention,
$K_\pm([0])!=1$. Hence, the general solution to (\ref{coh}) defines
states lying only within the subspace $\overline{\cal V}$, of the form
\begin{eqnarray}
\label{coh2}
|z;\tau_\pm;\theta,\phi\rangle=\sum_{n=0,\pm}^\infty\,
\left(\frac{q^{\mu}}{p^{\nu}}\right)^{n(n-1)/2}\,
\frac{z^n}{K_\pm([n])!}\,C^\pm_0(z)\,|E^\pm_n\rangle.
\end{eqnarray}
Note that the eigenvalue problem (\ref{coh}) is singular at the
particular value $z=0$, since its solution is an arbitrary superposition
of the three states $|E_*\rangle$ and $|E^\pm_0\rangle$. Nevertheless,
we shall consider the $(p,q)$-VCS associated to $z=0$,
$|z=0;\tau_\pm;\theta,\phi\rangle$, as being defined through the continuous
limit in $z\to 0$ of the construction in (\ref{coh2}), namely
$|z=0;\tau_\pm;\theta,\phi\rangle=C^+_0(0)|E^+_0\rangle +C^-_0(0)|E^-_0\rangle$.

Let us now turn to the issue of the stability of the $(p,q)$-VCS under
time evolution generated by the Hamiltonian (\ref{eq:hq}). Namely, we now
require furthermore that $(p,q)$-VCS are transformed into one another
under time evolution according to the following dependence on the real
parameters $\tau_\pm$, for all $t\in\mathbb{R}$,
\begin{eqnarray}
\label{evol}
e^{-i\omega_0 t\,{\cal H}^{\rm red}}\,|z; \tau_\pm;\theta,\phi\rangle
= |z; t+\tau_\pm;\theta,\phi\rangle.
\end{eqnarray}
Since one has, for all $n\in\mathbb{N}$,
\begin{equation}
e^{-i\omega_0 t\,{\cal H}^{\rm red}}\,|E^\pm_n\rangle=
e^{-i\omega_0 t\,E^\pm_n}\,|E^\pm_n\rangle,
\end{equation}
one needs to factor out their complex phases from the quantities $K_\pm([n])$,
\begin{eqnarray}
K_\pm([n]) = e^{i\varphi_\pm([n])}K^0_\pm([n]),
\end{eqnarray}
where $K^0_\pm([n])>0$ are now real positive scalars. The stability
condition (\ref{evol}) is then solved by choosing, for all $n=1,2,\cdots$,
\begin{equation}
\varphi_\pm([n])=\omega_0\tau_\pm\left[E^\pm_n-E^\pm_{n-1}\right],
\end{equation}
and redefining
\begin{equation}
C^\pm_0(z)={\cal C}^\pm_0(z)\,e^{-i\omega_0\tau_\pm E^\pm_0},
\end{equation}
where ${\cal C}^\pm_0(z)$ are new complex functions of $z$. Hence,
\begin{equation}
|z;\tau_\pm;\theta,\phi\rangle = \sum_{n=0,\pm}^\infty\,
\left(\frac{q^{\mu}}{p^{\nu}}\right)^{n(n-1)/2}\,
\frac{z^n}{K^0_\pm([n])!}\,{\cal C}^\pm_0(z)\,
e^{-i\omega_0\tau_\pm\,E^\pm_n}\,|E^\pm_n\rangle.
\label{coh3}
\end{equation}

Having identified both the $z$ and $\tau_\pm$ dependences
of the coherent states, finally let us account for their $(\theta,\phi)$
dependence and $S_2$ vector character implicit so far through the two functions
${\cal C}^\pm_0(z)$. The latter are now chosen to be given as
\begin{equation}
{\cal C}^+_0(z)=N^+(|z|)\,\cos\theta,\qquad
{\cal C}^-_0(z)=N^-(|z|)\,e^{i\phi}\,\sin\theta,
\end{equation}
$N^\pm(|z|)$ being factors such that the constructed $(p,q)$-VCS be of unit norm,
\begin{equation}
N^\pm(|z|)=\left\{\sum_{n=0}^\infty\,\left(\frac{q^{\mu}}{p^{\nu}}\right)^{n(n-1)}\,
\frac{|z|^{2n}}
{\left(K^0_\pm([n])!\right)^2}\right\}^{-1/2}.
\end{equation}
The convergence radii $R_\pm$ of these two series in $z$,
\begin{equation}
R_\pm=\lim_{n\to\infty}\left\{(q^{\mu}p^{-\nu})^{-(n-1)}\,K^0_\pm([n])\right\},
\end{equation}
depend on the choice of functions $K^0_\pm([n])$ as well
as on $(\mu,\nu)$ possibly. Specific cases are considered hereafter.

Consequently, the $(p,q)$-VCS constructed here are properly
defined provided $z\in D_R$ where $D_R$ denotes the disk in
the complex plane centered at $z=0$ and of radius
$R={\rm min}\left(R_+,R_-\right)$. Their general structure is thus
of the form
\begin{eqnarray}
\label{coh5}
|z;\tau_\pm;\theta,\phi\rangle &=& \ \ \ N^+(|z|)\,\cos\theta\,
\sum_{n=0}^\infty \left(\frac{q^{\mu}}{p^{\nu}}\right)^{n(n-1)/2}\,
\frac{z^n}{K^0_+([n])!}\,
e^{-i\omega_0\tau_+\,E^+_n}\,|E^+_n\rangle  \cr
&& +\ N^-(|z|)\,e^{i\phi}\,\sin\theta\,
\sum_{n=0}^\infty \left(\frac{q^{\mu}}{p^{\nu}}\right)^{n(n-1)/2}\,
\frac{z^{n}}{K^0_-([n])!}\,
e^{-i\omega_0\tau_-\,E^-_n}\,|E^-_n\rangle.
\end{eqnarray}
Only the real positive functions $K^0_\pm([n])$ still need to be
specified. They parametrize the remaining freedom in the construction.
Particular examples will be considered hereafter by imposing further
requirements on these $(p,q)$-VCS. Note that the double limit $p,q\to 1$
yields the VCS of the non-deformed ${\cal JC}m$
as obtained by Hussin and Nieto \cite{hus1}, briefly described in Section \ref{Sect2}.

\subsection{Some expectation values}
\label{Sect4.2}

Before dealing with further requirements on the family of $(p,q)$-VCS,
among which their overcompleteness in the space $\overline{\cal V}$,
let us consider some relevant expectation values for these states.
Given (\ref{coh5}), the mean value of ${\cal H}^{\rm red}$ for any
of the $(p,q)$-VCS is simply
\begin{eqnarray}
\label{min}
\langle{\cal H}^{\rm red}\rangle
&=&\ \ \ |N^+(|z|)|^2\,\cos^2\theta\,
\sum_{n=0}^\infty\, \left(\frac{q^{\mu}}{p^{\nu}}\right)^{n(n-1)}
\frac{|z|^{2n}}{\left(K^0_+([n])!\right)^2}\,E^+_n\cr
&& +\ |N^-(|z|)|^2\,\sin^2\theta\,
\sum_{n=0}^\infty\, \left(\frac{q^{\mu}}{p^{\nu}}\right)^{n(n-1)}
\frac{|z|^{2n}}{\left(K^0_-([n])!\right)^2}\,E^-_n.
\end{eqnarray}
Likewise for the ``number" operator associated to the ladder
operators ${\cal A}^-$ and ${\cal A}^+$, one finds the expectation value
\begin{eqnarray}
\langle{\cal A}^+\,{\cal A}^-\rangle
&=& |z|^{2}\left\{\ \ |N^+(|z|)|^2\,\cos^2\theta\,
\sum_{n=0}^\infty\, \left(\frac{q^{\mu}}{p^{\nu}}\right)^{n(n+1)}
\frac{|z|^{2n}}{\left(K^0_+([n])!\right)^2}\right.\,\cr
&&\ \ \ \ \ \ \ +\left. |N^-(|z|)|^2\,\sin^2\theta\,
\sum_{n=0}^\infty\, \left(\frac{q^{\mu}}{p^{\nu}}\right)^{n(n+1)}
\frac{|z|^{2n}}{\left(K^0_-([n])!\right)^2}\right\}.
\end{eqnarray}

Finally, the average atomic spin time evolution
$\langle\sigma_3(t)\rangle=\langle U^{-1}(t)\sigma_3 U(t)\rangle$, with
$U(t)=exp\{-i\omega_0 t\,{\cal H}^{\rm red}\}$ being the time evolution
operator, has the form
\begin{eqnarray}
&&\langle\sigma_3(t)\rangle =
\frac{1}{2}\sum_{n=0}^\infty  \left(\frac{q^{\mu}}{p^{\nu}}\right)^{n(n-1)}|z|^{2n}
\frac{{\cal E}([n+1])}{Q([n+1])}
\left\{-\frac{|N^+(|z|)|^2}{\left(K^0_+([n])!\right)^2}\cos^2\theta
+\frac{|N^-(|z|)|^2}{\left(K^0_-([n])!\right)^2}\sin^2\theta\right\}\cr
&&+ \lambda N^+(|z|)\,N^-(|z|)\sin 2\theta
\sum_{n=0}^\infty  \left(\frac{q^{\mu}}{p^{\nu}}\right)^{n(n-1)}
\frac{|z|^{2n}}{K^0_+([n])! K^0_-([n])!}
\frac{[n+1]}{Q([n+1])}\cos\Psi_n(t),
\label{ato}
\end{eqnarray}
with
\begin{equation}
\Psi_n(t)=\omega_0\left[\left(t+\tau_+\right)E^+_n\,-\,
\left(t+\tau_-\right)E^-_n\right]\,+\,\phi=
\omega_0\Delta E_n\,t\,+\,\omega_0\left[\tau_+ E^+_n - \tau_- E^-_n\right]+\phi.
\end{equation}
As is the case in the non-deformed model, the explicit time dependence which arises for
the atomic inversion $\langle\sigma_3(t)\rangle$ is due to the mixed state sector,
namely the fact that the mixed-spin matrix elements of the Heisenberg picture
operator $\sigma_3(t)$ do not vanish when $\lambda\ne 0$.
Hence, the proposition which states that the time dependence of atomic inversion
consists of Rabi oscillations when a system is prepared in a coherent state
of the radiation field \cite{hus4} extends to $(p,q)$-VCS. However, in the
limit where $\lambda\to 0$, no such oscillations occur.
Let us also point out that the time dependence of
$\langle\sigma_3(t)\rangle$ diplays chaotic behaviour for appropriate values
of the model parameters, as was previously mentioned for the $q$-deformation
of the model, with $0<q<1$, in the work by Naderi {\it et al.} \cite{nad}.

\subsection{Overcompleteness and the moment problem}
\label{Sect4.3}

An important property that coherent states ought to meet is that
of overcompleteness in the space over which they are defined \cite{kl}.
In the present case, this means that the $(p,q)$-VCS in (\ref{coh5})
must also provide a resolution of the identity operator over the
subspace $\overline{\cal V}$, namely
\begin{equation}
\mathbb{I}_{\cal V}=\mathbb{I}_{{\cal V}_0}\,+\,
\mathbb{I}_{\overline{\cal V}}=|E_*\rangle\,\langle E_*|\,+\,
\mathbb{I}_{\overline{\cal V}},
\end{equation}
while
\begin{equation}
\mathbb{I}_{\overline{\cal V}}=
\sum_{n=0,\pm}^\infty\,|E^\pm_n\rangle\,\langle E^\pm_n|=
\int_{D_R\times S_2}\,d\mu(z;\theta,\phi)\,
|z;\tau_\pm;\theta,\phi\rangle\,\langle z;\tau_\pm;\theta,\phi|,
\label{res}
\end{equation}
where $d\mu(z;\theta,\phi)$ is some SU(2) matrix-valued integration
measure over $D_R\times S_2$ to be determined from the above requirement.

Let us thus consider the following parametrization of that measure,
\begin{equation}
d\mu(z;\theta,\phi)=d^2z\,d\theta\,\sin\theta\,d\phi\,
\left\{{\cal W}^+(|z|)\sum_{n=0}^\infty|E^+_n\rangle\langle E^+_n|\,+\,
{\cal W}^-(|z|)\sum_{n=0}^\infty|E^-_n\rangle\langle E^-_n|\right\},
\label{eq:weight}
\end{equation}
in terms of real weight functions ${\cal W}^\pm(|z|)$ to be identified.
Using the radial parametrization $z=r\,e^{i\varphi}$ and $d^2z=dr\,r\,d\varphi$
where $r\in[0,\infty[$ and $\varphi\in[0,2\pi[$, a direct substitution in (\ref{res})
leads to the moment problem associated to the overcompleteness relation (\ref{res}).
In terms of the functions $h^\pm(r^2)$ defined through
\begin{equation}
h^+(r^2)=\frac{4\pi^2}{3}\,|N^+(r)|^2\,{\cal W}^+(r),\qquad
h^-(r^2)=\frac{8\pi^2}{3}\,|N^-(r)|^2\,{\cal W}^-(r),
\label{eq:hfunctions}
\end{equation}
the following two infinite sets of moment identities must be met, for all $n\in\mathbb{N}$,
\begin{equation}
\int_0^{R^2} du\,u^n\,h^\pm(u)= \left(\frac{q^{\mu}}{p^{\nu}}\right)^{-n(n-1)}\,
\left(K^0_\pm([n])!\right)^2.
\label{eq:moment}
\end{equation}

In conclusion, the resolution of the identity operator over $\overline{\cal V}$ in
terms of the $(p,q)$-VCS is achieved provided the Stieljes moment problem
(\ref{eq:moment}) can be solved \cite{hs,sim}. This requires a choice of
functions $K^0_\pm([n])>0$ such that not only the conditions (\ref{eq:moment})
may all be met, but also such that the normalization factors $N^\pm(|z|)$
converge in a non-empty disc of the complex plane.

As a result of this analysis, {\it a priori\/} there may exist a large number of sets
of $(p,q)$-VCS which fulfill all the above properties, namely continuity
in the complex parameter $z$, temporal stability through a simple additive
time dependence in the real parameters $\tau_\pm$, a unit vector valued characterization
on the sphere $S_2$ in terms of the spherical coordinates $\theta$ and $\phi$, and
the completeness property of a resolution of the unit operator with
a SU(2) matrix-valued integration measure over these spaces. These sets
of $(p,q)$-VCS are distinguished from one another by different choices of
real positive weight factors $K^0_\pm([n])$, in agreement with the considerations
developed in \cite{kl,kl2}. The above construction of $(p,q)$-VCS is general,
but can admit explicit exact solutions to the moment problem (\ref{eq:moment})
for particular cases. Concrete examples are discussed in Section \ref{Sect5}..

\subsection{Action-angle variables}
\label{Sect4.4}

One of the useful properties that general coherent states constructed
according to the arguments of \cite{kl2} possess, is that action-angle variables
are readily identified in relation to the continuous parameters ensuring
stability of the coherent states under time evolution. In the present
case, canonical reduced action-angle variables $(J_\pm(t),\tau_\pm(t))$ are such that
for the previously evaluated expectation values of the reduced Hamiltonian
(\ref{eq:hq}) in the $(p,q)$-VCS, one has
\begin{equation}
\langle{\cal H}^{\rm red}\rangle=J_+\,\omega_+\ +\ J_-\,\omega_-
=\sum_\pm\,J_\pm\,\omega_\pm,
\end{equation}
in relation to the action-angle variational principle of the form
\begin{equation}
\int\,dt\sum_\pm\,\left[\frac{d\tau_\pm}{dt}\,J_\pm\,-\,\omega_\pm\,J_\pm\right]\
\longleftrightarrow\
\int\,dt\left[\langle \frac{i}{\omega_0}\frac{d}{dt}\rangle\,-\,
\langle{\cal H}^{\rm red}\rangle\right] ,
\end{equation}
where $\omega_\pm$ are two constant factors to be chosen appropriately.
Consequently
\begin{equation}
\frac{d\tau_\pm}{dt}
=\frac{\partial\langle{\cal H}^{\rm red}\rangle}{\partial J_\pm}
=\omega_\pm,\qquad
\frac{dJ_\pm}{dt}
=-\frac{\partial\langle{\cal H}^{\rm red}\rangle}{\partial\tau_\pm}
=0.
\end{equation}
Given the time evolution, $\tau_\pm(t)=t+\tau_\pm(0)$,
one simply finds $\omega_\pm =1$. From the expression in (\ref{min}),
one then has the identifications
\begin{eqnarray}
J_+&=&|N^+(|z|)|^2\,\cos^2\theta\,
\sum_{n=0}^\infty\, \left(\frac{q^{\mu}}{p^{\nu}}\right)^{n(n-1)}\,\frac{|z|^{2n}}
{\left(K^0_+([n])!\right)^2}\,E^+_n, \cr
J_-&=&|N^-(|z|)|^2\,\sin^2\theta\,
\sum_{n=0}^\infty\, \left(\frac{q^{\mu}}{p^{\nu}}\right)^{n(n-1)}\,\frac{|z|^{2n}}
{\left(K^0_-([n])!\right)^2}\,E^-_n.
\label{eq:J}
\end{eqnarray}

As a final remark, let us mention that the saturated Heisenberg uncertainty
relations which are obeyed by $q$- and $(p,q)$-coherent states are also well-known
in $q$-mechanics (see for instance \cite{kemp}). Such minimal uncertainties may be
characterized through small corrections to canonical commutation relations defined
in \cite{kemp,pens}. Such properties in the case of the $(p,q)$-VCS
constructed here are deferred to a later study.

\section{Explicit Solutions}
\label{Sect5}

In order to completely specify the quantities $K^0_\pm([n])$, one last set
of conditions needs to be implemented. In the present Section, two such choices
are discussed, one of which allows for an exact and explicit solution to the moment problem,
hence the construction of a set of $(p,q)$-VCS. First, in line with the illustrative
example of Section \ref{Sect2}, we consider restricting the algebra of the ladder
operators ${\cal A}^\pm$. Then as a second and independent possibility, we apply
a final additional criterion developed in \cite{kl} in order to uniquely characterize a set
of coherent states which meet already all the requirements considered heretofore and having
led to the representation (\ref{coh5}), even though the moment problem remains unsolved for
that choice.

\subsection{Constraining the ladder operator algebra}
\label{Sect5.1}

In order to uniquely identify the set of functions $K^0_\pm([n])>0$,
let us consider the possibility that this may be achieved by restricting
the algebraic properties of the ladder operators. In line with the
general $(p,q)$-deformations of the Fock algebra in (\ref{eq:pq-def1}),
let us constrain the algebra of the operators $\mathbb{A}^\pm$ acting on
${\cal V}$ to be such that
\begin{eqnarray}
\mathbb{A}^-\,\mathbb{A}^+\,-\,q_0\,\mathbb{A}^+\,\mathbb{A}^-=p^{-N}_0
&=&\sum_{n=0,\pm}^\infty|n,\pm\rangle\,p^{-n}_0\,\langle n,\pm|,\cr
\mathbb{A}^-\,\mathbb{A}^+\,-\,p^{-1}_0\,\mathbb{A}^+\,\mathbb{A}^-=q^{N}_0
&=&\sum_{n=0,\pm}^\infty\,|n,\pm\rangle\,q^n_0\,\langle n,\pm|,
\label{eq:const1}
\end{eqnarray}
where $p_0$ and $q_0$ are again two real parameters such that $p_0>1$, $0<q_0<1$
and $p_0 q_0<1$, which may or may not be identical to $p$ and $q$. For instance,
we could have $p_0=1$ and $q_0=1$ thus corresponding to an ordinary Fock algebra,
or else $p_0=p$ and $q_0=q$, but also more generally $p_0=p^\alpha$ and $q_0=q^\alpha$,
$\alpha$ being some real constant. As a matter of fact, exact solutions
to the moment problem are presented hereafter in all these situations.

In terms of the ladder operators ${\cal A}^\pm={\cal U}\,\mathbb{A}^\pm\,{\cal U}^\dagger$
acting on the subspace $\overline{\cal V}$, the associated algebraic constraint reads
\begin{eqnarray}
{\cal A}^-\,{\cal A}^+\,-\,q_0\,{\cal A}^+\,{\cal A}^-
&=&\sum_{n=0,\pm}^\infty|E^\pm_n\rangle\,p^{-n}_0\,\langle E^\pm_n|,\cr
{\cal A}^-\,{\cal A}^+\,-\,p^{-1}_0\,{\cal A}^+\,{\cal A}^-
&=&\sum_{n=0,\pm}^\infty\,|E^\pm_n\rangle\,q^n_0\,\langle E^\pm_n|.
\label{eq:const2}
\end{eqnarray}
Whether in terms of (\ref{eq:const1}) or (\ref{eq:const2}), these algebraic constraints
translate into the following identities, for all $n\in\mathbb{N}$,
\begin{equation}\label{eq:recrel}
\left(K^0_\pm([n+1])\right)^2\,-\,q_0\,\left(K^0_\pm([n])\right)^2=p^{-n}_0,\quad
\left(K^0_\pm([n+1])\right)^2\,-\,p^{-1}_0\,\left(K^0_\pm([n])\right)^2=q^n_0.
\end{equation}
Given the initial values $K^0_\pm([0])=0$, the solution to these recursion relations is simply
\begin{equation}
K^0_\pm([n])=\sqrt{[n]_{(p_0,q_0)}}=
\sqrt{[n]_{(q^{-1}_0,p^{-1}_0)}},
\label{eq:sol1K0}
\end{equation}
where\footnote{Incidentally, it is because of this identity, corresponding to the
exchange $p_0\leftrightarrow q^{-1}_0$, that the two solutions
to the above two recursion relations are consistent, as are the two algebraic
restrictions in (\ref{eq:const1}) and (\ref{eq:const2}).}
\begin{equation}
[n]_{(p_0,q_0)}=\frac{p^{-n}_0-q^n_0}{p^{-1}_0-q_0}=
\frac{\left(q^{-1}_0\right)^{-n}-\left(p^{-1}_0\right)^n}
{\left(q^{-1}_0\right)^{-1}-\left(p^{-1}_0\right)}=[n]_{(q^{-1}_0,p^{-1}_0)}.
\end{equation}

Given this solution, the normalization factors are defined by the series
\begin{equation}
|N^\pm(|z|)|^{-2}=\sum_{n=0}^\infty\,\left(\frac{q^{\mu}}{p^{\nu}}\right)^{n(n-1)}\,
\frac{|z|^{2n}}{[n]_{(p_0,q_0)}!},
\end{equation}
of which the convergence radius is
\begin{equation}
R=\lim_{n\to\infty}\left[\left(\frac{q^{\mu}}{p^{\nu}}\right)^{-2(n-1)}
\frac{p^{-n}_0-q^n_0}{p^{-1}_0-q_0}\right]^{1/2}
=\lim_{n\to\infty}
\left[\left(p_0p^{-2\nu}q^{2\mu}\right)^{-(n-1)}\,\frac{1-(p_0q_0)^n}{1-(p_0q_0)}\right]^{1/2}.
\end{equation}
Provided $p_0p^{-2\nu}q^{2\mu}<1$, a condition which we shall henceforth assume to be
satisfied\footnote{If $p_0p^{-2\nu}q^{2\mu}=1$, the radius of convergence is finite
with $R=(1-p_0q_0)^{-1/2}$, while when $p_0p^{-2\nu}q^{2\mu}>1$ the radius of
convergence vanishes, implying that $(p,q)$-VCS cannot be constructed in such a case.},
this radius of convergence is infinite, $R=\infty$, and the moment problem (\ref{eq:moment})
then becomes, for all $n\in\mathbb{N}$,
\begin{equation}
\int_0^\infty\,du\,u^n\,h^\pm(u)=\left(\frac{q^{\mu}}{p^{\nu}}\right)^{-n(n-1)}\,
\left([n]_{(p_0,q_0)}!\right).
\label{eq:moment2}
\end{equation}
In order to solve these equations, the Ramanujan integral (\ref{eq:App-Ramanujan})
discussed in the Appendix suggests itself quite naturally, through a simple but
appropriate rescaling of its arguments in the form of (\ref{eq:App-Ramanujan2}).

After a little moment's thought one comes to the conclusion that a solution
to (\ref{eq:moment2}) based on (\ref{eq:App-Ramanujan2}) is possible
for the following choice of parameters,
\begin{equation}
\mu=\frac{1}{2},\qquad \nu=0,\qquad p_0=p,\qquad q_0=q,
\end{equation}
in which case $p_0 p^{-2\nu} q^{2\mu}=pq<1$, hence corresponding indeed to
an infinite radius of convergence. For this choice, one has (for definitions
of the $(p,q)$-exponential functions appearing in these expressions, see
the Appendix),
\begin{equation}
h^\pm\left(|z|^2\right)=\frac{\left(p^{-1}-q\right)}{q\log\left(1/pq\right)}\,
e_{(p,q)}\left(-|z|^2\,p^{-1/2} q^{-1}\left(p^{-1}-q\right)\right),
\label{eq:sol-mom}
\end{equation}
as well as\footnote{Restricting to $p_0=p$ and $q_0=q$ but keeping $\mu$ and $\nu$ arbitrary
such that $p^{1-2\nu} q^{2\mu}<1$ in order to retain an infinite radius of convergence, one has
$\left(K^0_\pm([n])\right)^2=[n]$ and
$|N^\pm\left(|z|\right)|^{-2}={\cal E}^{(\mu,\nu)}_{(p,q)}
\left(|z|^2\,p^\nu\,q^{-\mu}\left(p^{-1}-q\right)\right)$, hence also all
other previous expressions given accordingly.}
\begin{equation}
\left(K^0_\pm([n])\right)^2=[n],\qquad
|N^\pm(|z|)|^{-2}
={\cal E}^{(1/2,0)}_{(p,q)}\left(|z|^2q^{-1/2}\left(p^{-1}-q\right)\right),
\end{equation}
with for the weight functions ${\cal W}^\pm(|z|)$ in the integration measure (\ref{eq:weight})
of the overcompleteness relation (\ref{res}),
\begin{equation}
{\cal W}^+\left(|z|\right)=\frac{3}{4\pi^2}\,|N^+\left(|z|\right)|^{-2}\,h^+\left(|z|^2\right),\qquad
{\cal W}^-\left(|z|\right)=\frac{3}{8\pi^2}\,|N^-\left(|z|\right)|^{-2}\,h^-\left(|z|^2\right).
\label{eq:weight2}
\end{equation}
Explicit expressions for all previously computed quantities readily follow, beginning
with the definition of the associated $(p,q)$-VCS which then meet all the necessary requirements
expected of coherent states. Note that up to the coefficients
$3/(2\pi)$ and $3/(4\pi)$, the reduced weights obtained are compatible with
that of the $q$-shape invariant harmonic oscillator \cite{bal2}.
Furthermore, (\ref{eq:sol-mom}) is a $(p,q)$-generalization of the
$q$-harmonic oscillator coherent state moment problem solution constructed
in \cite{quesne}. Finally, in the double limit $p,q\to 1$, the results of \cite{hus1}
are recovered.

The functions (\ref{eq:sol-mom}) thus provide a complete and explicit solution to the moment problem
of the $(p,q)$-VCS for the $(p,q)$-${\cal JC}m$ such that the ladder operators ${\cal A}^\pm$ obey
the same $(p,q)$-Fock algebra as the original modes $a$ and $a^\dagger$ of the initial Hamiltonian (\ref{eq:hq}),
namely with the choice $p_0=p$ and $q_0=q$. It is also possible to construct an explicit solution
when the ladder operators ${\cal A}^\pm$ are constrained to rather obey the ordinary non-deformed Fock
algebra on $\overline{\cal V}$, corresponding to the choice $p_0=1$ and $q_0=1$. One then has
to consider\footnote{Leading to $|N^\pm\left(|z|\right)|^{-2}=
\mathfrak{e}^{(\mu,\nu)}_{(p,q)}\left(|z|^2\,p^\nu\,q^{-\mu}\right)$, which converges for all $|z|<\infty$
provided $p^{-\nu}q^\mu\le 1$.}, for all $n\in\mathbb{N}$,
\begin{equation}
K^0_\pm([n])=\sqrt{n},\qquad
\int_0^\infty\,du\,u^n\,h^\pm(u)=
\left(\frac{q^\mu}{p^\nu}\right)^{-n(n-1)}\,\left(n!\right),\qquad
p^{-\nu}q^\mu\le 1.
\end{equation}
An obvious solution to this moment problem is obtained when $\mu=0=\nu$,
in which case the condition for an infinite radius of convergence is saturated.
One then has
\begin{equation}
h^\pm\left(|z|^2\right)=e^{-|z|^2},\qquad
|N^\pm\left(|z|\right)|^{-2}=e^{|z|^2},\qquad
{\cal W}^+\left(|z|\right)=\frac{3}{4\pi^2},\qquad
{\cal W}^-\left(|z|\right)=\frac{3}{8\pi^2}.
\end{equation}

In fact, the above two explicit solutions belong to a general class of
solutions obtained by taking $(p_0,q_0)=(p^\alpha,q^\alpha)$ with $\alpha$ a positive
real parameter, $\alpha>0$, such that $p^{\alpha-2\nu}q^{2\mu}<1$ in order to ensure
an infinite radius of convergence\footnote{Leading to $|N^\pm\left(|z|\right)|^{-2}=
{\cal E}^{(\mu/\alpha,\nu/\alpha)}_{(p^\alpha,q^\alpha)}\left(|z|^2p^\nu q^{-\mu}(p^{-\alpha}-q^\alpha)\right)$.}
in $z\in\mathbb{C}$. Once again based on (\ref{eq:App-Ramanujan2}),
an explicit solution to the moment problem (\ref{eq:moment2}) is achieved for the following choice
of parameters,
\begin{equation}
\mu=\frac{1}{2}\alpha,\qquad \nu=0,\qquad p_0=p^\alpha,\qquad q_0=q^\alpha,
\end{equation}
for which the radius of convergence is indeed infinite,
$p^{\alpha-2\nu}q^{2\mu}=(pq)^\alpha<1$. One then has
\begin{equation}
h^\pm\left(|z|^2\right)=\frac{\left(p^{-\alpha}-q^\alpha\right)}
{q^\alpha\log\left(1/p^\alpha q^\alpha\right)}\,
e_{(p^\alpha,q^\alpha)}\left(-|z|^2\,p^{-\alpha/2} q^{-\alpha}\left(p^{-\alpha}-q^\alpha\right)\right),
\label{eq:sol-momalpha}
\end{equation}
with
\begin{equation}
|N^\pm(|z|)|^{-2}
={\cal E}^{(1/2,0)}_{(p^\alpha,q^\alpha)}\left(|z|^2q^{-\alpha/2}\left(p^{-\alpha}-q^\alpha\right)\right),
\end{equation}
leading finally to the weight functions ${\cal W}^\pm(|z|)$ given in terms of the latter two quantities
through the same relations as in (\ref{eq:weight2}). In the limits that $\alpha\to 1$ or $\alpha\to 0$,
the previous two explicit solutions are then recovered as particular cases.

\subsection{The action identity constraint}
\label{Sect5.2}

An alternative to fixing the factors $K^0_\pm([n])$ through conditions on the
algebra of ladder operators, is to consider the action identity constraint
discussed in \cite{kl} as the one last requirement which singles out
coherent states uniquely. In the case of the ordinary Fock algebra,
this action identity constraint is equivalent to requiring that the ladder
operators obey themselves the Fock algebra as well. We shall establish
that this is not the case for the $(p,q)$-VCS of the $(p,q)$-${\cal JC}m$
constructed above.

Given the relations (\ref{eq:J}), in the present model the action identity
constraint is of the form
\begin{equation}
J_+=\cos^2\theta \left(|z|^2 + E^+_0\right),\qquad
J_-=\sin^2\theta \left(|z|^2 + E^-_0\right).
\label{eq:J2}
\end{equation}
By direct substitution into these constraints of the relations (\ref{eq:J}),
the identification of the successive powers in $|z|^2$ leads to the following
solution for the factors $K^0_\pm([n])$,
\begin{equation}
K^0_\pm([n])=\left(\frac{q^\mu}{p^\nu}\right)^{(n-1)}\,
\sqrt{E^\pm_n\,-\,E^\pm_0}.
\end{equation}
These positive real quantities are thus well-defined provided
one has $E^\pm_n>E^\pm_0$ for all $n\ge 1$, as is implicitly assumed.
It is noteworthy that, as $(p,q)\to (1^+,1^-)$,  these factors reduce to exactly
those obtained in \cite{hus3} by the factorization method. On the other hand,
since the present solution for $K^0_\pm([n])$ cannot be brought into the form
of (\ref{eq:sol1K0}) for some choice of constants $p_0$ and $q_0$ meeting our
assumptions for these quantities, it follows indeed that for the $(p,q)$-${\cal JC}m$
the action identity constraint is not equivalent to requiring an algebraic constraint
on the ladder operators of the $(p_0,q_0)$-deformed Fock algebra type.

This choice also allows for the factorization of the Hamiltonian in (\ref{eq:hpq})
in the form
\begin{equation}
\mathbb{H}^{\rm red}=\mathbb{A}^+\,\left(\frac{q^\mu}{p^\nu}\right)^{-2N}\,\mathbb{A}
\ +\ \sum_{n=0,\pm}^\infty\,|n,\pm\rangle\,E^\pm_0\,\langle n,\pm|,
\end{equation}
extending a similar expression in \cite{hus1}.

Given this solution for the factors $K^0_\pm([n])$, the general moment problem
(\ref{eq:moment}) reduces to the following conditions,
\begin{equation}
\int_0^{R^2}du\,\,h^\pm(u)=1;\qquad
\int_0^{R^2}du\,u^n\,h^\pm(u)=
\prod_{k=1}^n \left(E^\pm_k-E^\pm_0\right),\quad n=1,2,3,\cdots,
\end{equation}
where the radius of convergence $R$ is given as
\begin{equation}
R={\rm min}\,\left(R_+,R_-\right),\qquad
R_\pm=\lim_{n\to+\infty}\sqrt{E^\pm_n-E^\pm_0}.
\end{equation}
In the absence of a detailed analysis of the energy spectra $E^\pm_n$ as functions
of the parameters $p$, $q$, $\lambda$ and $\epsilon$ and the function $h(p,q)$,
nothing more explicit may be said concerning this moment problem. Since when
$p>1$ the quantities $[n]$ always possess a turn-around behaviour as functions of
$n$ for $n$ sufficiently large, it is to be expected generally that the radius of
convergence $R$, hence the moment problem as well, are associated to a finite disk
$D_R$ in the complex plane. Nevertheless, one conclusion of the present discussion
is that indeed for the $(p,q)$-VCS considered in this work, the action identity constraint
leads to coherent states different from those constructed in Section \ref{Sect5.1} and
for which explicit solutions to the moment problem have been given.

\subsection{The spin decoupled limit $\lambda=0$}
\label{Sect5.3}

In the limit that $\lambda=0$, the two spin sectors of
the model are decoupled, and the $(p,q)$-${\cal JC}m$
reduces to the supersymmetric harmonic oscillator \cite{arag,org,daou}
with a $(p,q)$-deformation. Diagonalization of the reduced Hamiltonian
(\ref{eq:hq}) is then of course straightforward in the $\sigma_3$-eigenbasis, with,
for $n=0,1,2,\cdots$,
\begin{equation}
{\cal H}^{\rm red}_{\lambda=0}\,|n,\pm\rangle=\epsilon^\pm_n\,|n,\pm\rangle,\qquad
\epsilon^\pm_n=(1+\epsilon)h(p,q)[n]\,+\,\frac{1}{2}(1+\epsilon)\pm\frac{1}{2}.
\label{eq:lambda0}
\end{equation}

From that point of view, one thus has two decoupled $(p,q)$-deformed Fock bases,
for which one could consider the usual $(p,q)$-coherent states in each spin sector
separately. However, such coherent states do not coincide with any of those
constructed in this paper and obtained in the limit $\lambda=0$, because of the
distinguished role played by the singleton state $|E_*\rangle=|0,-\rangle$
and the $S_2$ unit vector character of the $(p,q)$-VCS.
In particular the ladder operators ${\cal A}^\pm$ acting within each of the
towers $|E^\pm_n\rangle$ do not coincide with the annihilation and creation
operators $a$ and $a^\dagger$ defining the Hamiltonian (\ref{eq:hq}), even in
the decoupled limit $\lambda=0$. As a matter of fact, the action of the
ladder operators ${\cal A}^\pm$ may switch between the two spin sectors
as a function of $n$ depending on the sign of the quantity ${\cal E}([n+1])$.

More specifically, let us introduce the notation
\begin{equation}
s_n={\rm sign}\,{\cal E}([n+1]),\qquad n\in\mathbb{N}.
\end{equation}
In the limit that $\lambda=0$, one has $Q([n+1])=|{\cal E}([n+1])|/2$, so that
the mixing angle $\theta([n])$ is now such that, for all $n\in\mathbb{N}$,
\begin{equation}
\lambda=0:\quad
\sin\theta([n])=\frac{1}{2}(1-s_n)\,({\rm sign}\,\lambda),\quad
\cos\theta([n])=\frac{1}{2}(1+s_n).
\end{equation}
Consequently, the towers of energy eigenstates $|E^\pm_n\rangle$ are then given as
follows, for all $n\in\mathbb{N}$,
\begin{equation}
\begin{array}{rll}
{\rm If}\ s_n=+1: &\quad |E^+_n\rangle_{\lambda=0}=|n+1,-\rangle,\quad
&|E^-_n\rangle_{\lambda=0}=|n,+\rangle ;\\
 & & \\
{\rm If}\ s_n=-1: &\quad
|E^+_n\rangle_{\lambda=0}=({\rm sign}\,\lambda)\,|n,+\rangle,\quad
&|E^-_n\rangle_{\lambda=0}=-({\rm sign}\,\lambda)\,|n+1,-\rangle,
\end{array}
\end{equation}
while the energy eigenvalues are given as
\begin{equation}
\begin{array}{rl}
{\rm If}\ s_n=+1: &\quad
E^+_n(\lambda=0)=(1+\epsilon)h(p,q)[n+1]\,+\,\frac{1}{2}(1+\epsilon)\,-\,\frac{1}{2}, \\
 & \\
&\quad E^-_n(\lambda=0)=(1+\epsilon)h(p,q)[n]\,+\,\frac{1}{2}(1+\epsilon)\,+\,\frac{1}{2}; \\
 &  \\
{\rm If}\ s_n=-1: &\quad
E^+_n(\lambda=0)=(1+\epsilon)h(p,q)[n]\,+\,\frac{1}{2}(1+\epsilon)\,+\,\frac{1}{2}, \\
 & \\
&\quad E^-_n(\lambda=0)=(1+\epsilon)h(p,q)[n+1]\,+\,\frac{1}{2}(1+\epsilon)\,-\,\frac{1}{2}.
\end{array}
\end{equation}
These spectra do indeed coincide with those in (\ref{eq:lambda0}), once the singleton
state $|E_*\rangle=|0,-\rangle$ with $E_*=\epsilon/2$ is included as well.

These expressions show how, even in the decoupled spin limit $\lambda=0$,
the $(p,q)$-VCS constructed here are not simply the juxtaposition of two
separate $(p,q)$-coherent states of the $(p,q)$-deformed Fock algebra in
each of the two spin sectors. Since the spectrum of the system is discrete
infinite, by leaving aside the singleton state $|0,-\rangle$, all the remaining
states still allow for similar types of constructions of coherent states,
but in such a way that different spin sectors are getting superposed,
leading to the SU(2) vector coherent states of the type studied here.
All the expressions detailed in the previous sections
for the $(p,q)$-VCS may readily be particularized to the limit $\lambda\to 0$.

\section{Conclusion}
\label{Sect6}

In this work, we considered $(p,q)$-deformations
of the Jaynes-Cummings model in the rotating wave approximation,
extending recent developments on this topic in the non-deformed
case \cite{hus1}. Having introduced $(p,q)$-deformed
versions of the model, first its energy eigenspectrum has been
identified, enabling the definition of different relevant
operators acting on Hilbert space and the characterization of
the spectrum in terms of two separate infinite discrete towers and
a singleton state. Among these operators, ladder operators
acting within each of the two towers separately may be considered,
defined up to some arbitrary normalization factors.

Such a structure sets the stage for the introduction of
vector coherent states for the $(p,q)$-deformed Jaynes-Cummings model,
following the approach of \cite{hus1} and the rationale outlined
in \cite{kl}. These $(p,q)$-VCS are parametrized by elements of
$\mathbb{C}\times S_2$, and enjoy temporal stability through
a further action-angle identification. The moment problem associated
to the overcompleteness property of these $(p,q)$-VCS involves
SU(2)-valued matrix weight functions. Using $(p,q)$-arithmetic
techniques, some explicit and exact solutions to the moment
problem have been displayed, hence characterizing specific classes
of such $(p,q)$-VCS. All these solutions provide $(p,q)$-extensions
to the non-deformed vector coherent states of the ${\cal JC}m$
considered in \cite{hus1}. These explicit solutions are
obtained by requiring that specific algebraic constraints of the
$(p,q)$-deformed Fock algebra type be obeyed by the ladder operators.
However, in contradistinction to \cite{hus1}, we have not been able
to display an explicit and exact solution to the moment problem in the
generic case by imposing an action identity constraint.

Finally, the spin decoupled limit of these models was considered,
corresponding to a $(p,q)$-supersymmetric oscillator
of which the two sectors are intertwined in a manner depending
on the sign of the energy level spacing between the two decoupled
spin sectors as function of the excitation level. In the non-deformed
limit $(p,q)=(1,1)$, this feature disappears, reproducing the
ordinary supersymmetric oscillator. Our results thus provide
new classes of generalized versions of the ${\cal JC}m$
in the rotating wave approximation \cite{bal2,daou}.
Finally, the $(p,q)$-VCS built here extend
the $q$-coherent states obtained by other techniques
involving supersymmetric shape invariance
and self-similar potential formalisms
applied to the harmonic oscillator \cite{bal2,coo}.

\section*{Acknowledgements}

J.~B.~G. is grateful to the Abdus Salam International Centre for Theoretical
Physics (ICTP, Trieste, Italy) for a Ph.D. fellowship under the
grant \mbox{Prj-15}. M.~N.~H. is particularly indebted to V. Hussin
for discussions relating to the ${\cal JC}m$ as well as for provided references during
his stay at the Centre de Recherches Math\'ematiques, Universit\'e de Montr\'eal,
Canada. The ICMPA is in partnership with the Daniel Iagoniltzer Foundation (DIF), France.

J.~G. acknowledges a visiting appointment as Visiting Professor in the School of
Physics (Faculty of Science) at the University of New South Wales. He is grateful
to Prof. Chris Hamer and the School of Physics for their hospitality during his sabbatical
leave, and for financial support through a Fellowship of the Gordon Godfrey Fund.
His stay in Australia is also supported in part by the Belgian National
Fund for Scientific Research (F.N.R.S.) through a travel grant.

J.~G. acknowledges the Abdus Salam International Centre for Theoretical
Physics (ICTP, Trieste, Italy) Visiting Scholar Programme
in support of a Visiting Professorship at the ICMPA. His work
is also supported by the Belgian Federal Office for Scientific,
Technical and Cultural Affairs through the Interuniversity Attraction
Pole (IAP) P5/27.

\section*{Appendix}
\label{App}

This appendix lists some useful facts related to the
$(p,q)$-boson algebra and associated functions.
The $(p,q)$-deformed oscillator algebra introduced in \cite{cha}
is generated by operators $a$, $a^{\dag}$ and $N$ obeying
the relations
\begin{eqnarray}
\label{pqalg}
& [N,a]=-a,&\quad [N,a^\dagger]=a^\dagger,\cr
& aa^\dagger-qa^\dagger a=p^{-N},&\quad
aa^\dagger-p^{-1}a^\dagger a=q^{N}.
\end{eqnarray}
Throughout the text, we assume the real parameters $p$ and $q$
are such that $p>1$, $0<q<1$ and $pq<1$. The limit $p\to 1^{+}$ yields
the $q$-oscillator of Arik and Coon \cite{ar} while $p=q$ gives
the $q$-deformed oscillator algebra of Biedenharn and MacFarlane \cite{far}.
Finally, the algebra (\ref{pqalg}) reduces to the ordinary
harmonic oscillator Fock algebra as $q\to 1$ for $p=1^{+}$ or $p=q$.
At any stage of the discussion, the $(p,q)$-deformed model readily
reduces to its usual counterpart as $(p,q) \to (1,1)$.

The associated $(p,q)$-deformed Fock-Hilbert space representation
is spanned by the vacuum $|0\rangle$ annihilated by $a$ and the orthonormalized
states $|n\rangle$, such that
\begin{eqnarray}
\label{pqrep}
&&
a|0\rangle=0,\quad \langle 0|0\rangle=1,\quad
|n\rangle = \frac{1}{\sqrt{[n]_{(p,q)}!}}\left(a^\dagger\right)^n|0\rangle,\cr
&&
a|n\rangle=\sqrt{[n]_{(p,q)}}|n-1\rangle,\quad
a^\dagger |n\rangle = \sqrt{[n+1]_{(p,q)}}|n+1\rangle,\quad
N|n\rangle  =n|n\rangle,
\end{eqnarray}
where the symbol $[n]_{(p,q)}=\left(p^{-n}-q^{n}\right)/\left(p^{-1}-q\right)$
is called $(p,q)$-basic number with, by convention, $[0]_{(p,q)}=0$,
and its $(p,q)$-factorial is defined through
$[n]_{(p,q)}!=[n]_{(p,q)}\left([n-1]_{(p,q)}!\right)$
and the convention $[0]_{(p,q)}!=1$. There exists a formal $(p,q)$-number operator
denoted by $[N]_{(p,q)}$, or simply by $[N]$ when no confusion arises.
As a matter of fact, from the second pair of relations in (\ref{pqalg}),
it follows that $[N]=a^\dagger a$ as well as $[N+1]=a a^\dagger$.
One has of course $[N]|n\rangle= [n]|n\rangle$. Hence, (\ref{pqrep})
provides a well defined Fock-Hilbert representation space of the
algebra (\ref{pqalg}).

The following relations hold for any function $f\equiv f(N)$
and consequently for any function of $[N]$,
\begin{eqnarray}
\label{afa}
a f(N -1)= f(N)a,\quad a^\dagger f(N) = f(N -1)a^\dagger.
\end{eqnarray}

Let us define $q$-shifted products and factorials
and their $(p,q)$-analogues. Using the notations of \cite{ga},
for any quantity $x$, $(x;q)_{\alpha}$ is constructed as follows,
\begin{eqnarray}
(x;q)_0=1,\qquad(x;q)_\alpha=\frac{(x;q)_\infty}{(xq^\alpha;q)_\infty},
\qquad (x;q)_\infty= \prod_{n=0}^\infty\left(1-xq^n \right).
\end{eqnarray}
Furthermore, in the notations of \cite{vin2},
$(p,q)$-shifted products and factorials
are defined as follows, for any real quantities $a$ and $b$
such that $a\neq 0$,
\begin{eqnarray}
[a,b;p,q]_{0}=1,\qquad
[a,b;p,q]_\alpha=\frac{[a,b;p,q]_\infty}{[ap^\alpha,bq^\alpha;p,q]_\infty},\qquad
[a,b;p,q]_\infty= \prod_{n=0}^\infty\left(\frac{1}{ap^n}-bq^n \right).
\end{eqnarray}
For $\alpha=n \in \mathbb{N}$, we have
\begin{eqnarray}
\label{pqprod}
[p^\mu,q^\nu;p,q]_n&=&\left(\frac{1}{p^\mu}-q^\nu \right)
\left(\frac{1}{p^{\mu+1}}-q^{\nu+1}\right)\dots
\left(\frac{1}{p^{\mu+n-1}}-q^{\nu+n-1}\right)\cr
&=&p^{-\mu n - n(n-1)/2}(p^\mu q^\nu;pq)_n.
\end{eqnarray}
This identity is a central formula since it defines a bridge
between $q$- and $(p,q)$-analogue quantities and functions.

Let us now introduce $q$-analogues of the ordinary exponential funtion.
There exist many types of $q$-deformations of the exponential function
$e^{z}$, $z\in \mathbb{C}$ (see, for instance, \cite{vin}).
For any $(z,\mu) \in \mathbb{C}\times\mathbb{R}$,
the $(\mu,q)$-exponential is the complex function \cite{vin}
\begin{eqnarray}
\label{muex}
E_{q}^{(\mu)}(z)= \sum_{n=0}^{\infty}\frac{q^{\mu n^{2}}}
{(q;q)_{n}}z^{n}.
\end{eqnarray}
This series has an infinite radius of convergence for $\mu>0$.
For $\mu=0$ its domain of definition reduces to the unit disk, $|z|<1$,
while it is nowhere convergent in $\mathbb{C}$ for $\mu<0$.
Rescaling $z \to z(1-q)$ and taking the limit $\lim_{q\to 1}E_{q}^{\mu}(z(1-q))$,
one recovers $e^{z}$. For some specific values of $\mu$, (\ref{muex})
reproduces some standard $q$-exponentials \cite{vin,koe},
\begin{eqnarray}
\label{eq}
E_{q}^{(0)}(z)&=& e_{q}(z)=\frac{1}{(z;q)_{\infty}}=
\sum_{n=0}^{\infty}\frac{z^{n}}{(q;q)_{n}},\qquad |z|<1,\\
\label{edemi}
E_{q}^{(1/2)}(z) &=& E_{q}(q^{1/2}z)=(-q^{1/2}z;q)_{\infty},
\qquad z\in \mathbb{C},
\end{eqnarray}
where
\begin{eqnarray}
\label{eqjac}
E_{q}(z)= \sum_{n=0}^{\infty}\frac{q^{n(n-1)/2} z^{n}}{(q;q)_{n}},
\qquad z\in\mathbb{C},
\end{eqnarray}
is known as the Jackson $q$-exponential \cite{jac}.
Note that whereas $E_{q}^{(\mu)}(z)$ is defined in the entire complex
plane, $|z|<\infty$, for any $\mu > 0 $, its reduction $e_{q}(z)$
is only defined on the unit disc. Finally, it is also well established
that \cite{koe}
\begin{eqnarray}
\label{invert}
E_{q}(-z)e_{q}(z)=1.
\end{eqnarray}

$(p,q)$-analogues of the usual exponential function $e^{z}$,
$z\in \mathbb{C}$ may also be introduced (see, for instance, \cite{vin2}).
Given any $(z,\mu,\nu) \in \mathbb{C}\times\mathbb{R}\times\mathbb{R}$,
consider the $(\mu,\nu,p,q)$-exponential function
\begin{eqnarray}
\label{pgex}
{\cal E}_{(p,q)}^{(\mu,\nu)}(z)= \sum_{n=0}^{\infty}
\left( \frac{q^{\mu}}{p^{\nu}} \right)^{n^{2}}
\frac{z^{n}}{[p,q;p,q]_{n}}.
\end{eqnarray}
Keeping in mind the condition $pq < 1$, the radius of convergence $R$
of this series is such that
\begin{eqnarray}
\label{rad}
R_1= \left\{\begin{array}{ll}
\infty, & \qquad {\rm if}\ q^{2\mu} p^{1-2\nu} < 1;  \\
p^{\nu-1}q^{-\mu}, & \qquad {\rm if}\ q^{2\mu}p^{1-2\nu} = 1; \\
0,  & \qquad {\rm if}\ q^{2\mu}p^{1-2\nu} > 1.
\end{array}\right.
\end{eqnarray}
Thus the function ${\cal E}^{(\mu,\nu)}_{(p,q)}(z)$ exists only provided
$q^{2\mu}p^{1-2\nu}\le 1$.

In order to recover the usual exponential function, one has to rescale $z \to z(p^{-1}-q)$,
for example, and then take the limit
$\lim_{(p,q)\to (1,1)}{\cal E}_{(p,q)}^{\mu,\nu}(z(p^{-1}-q))=e^{z}$.
For particular values of the parameters $\mu$ and $\nu$,
(\ref{pgex}) reproduces known $(p,q)$-exponentials,
\begin{eqnarray}
\label{pgex2}
{\cal E}_{(p,q)}^{(1/2,1/2)}(z)&=&
 E_{(p,q)}\left(\left(\frac{q}{p}\right)^{1/2}z\right)=
\sum_{n=0}^{\infty} \left(\frac{q}{p}\right)^{n^{2}/2}
\frac{z^{n}}{[p,q;p,q]_{n}},
\end{eqnarray}
where
\begin{eqnarray}
\label{epq}
E_{(p,q)}(z)&=&\sum_{n=0}^{\infty}
\left(\frac{q}{p}\right)^{n(n-1)/2}
\frac{z^{n}}{[p,q;p,q]_{n}}.
\end{eqnarray}
The function $E_{(p,q)}$ may be found in \cite{vin2}.
Note that (\ref{epq}) coincides with (\ref{eqjac}) as $p\to 1$.
In the same limit, (\ref{pgex}) reproduces the
$(\mu,q)$-deformed exponential map $E^{(\mu)}_q(z)$ \cite{vin}.
If $\mu=0=\nu$ the series (\ref{pgex}) is not defined since
then $R=0$, unless one has taken $p=1$ in which case the radius
of convergence is unity. A $(p,q)$-analogue of (\ref{eq}) is given by
\begin{eqnarray}
\label{egex}
e_{(p,q)}(z) =\sum_{n=0}^{\infty}
\frac{1}{p^{n^{2}/2}}\frac{z^{n}}{[p,q;p,q]_{n}},\qquad |z|<p^{-1/2},
\end{eqnarray}
which reproduces exactly $e_{q}(z)$ converging in the unit disc
as $p\to 1^{+}$. Furthermore, we have from (\ref{pqprod})
\begin{eqnarray}
\label{inv}
e_{(p,q)}(z)=\sum_{n=0}^{\infty}
\frac{(p^{1/2}z)^{n}}{(pq;pq)_{n}} =
 e_{pq}(p^{1/2}z) .
\end{eqnarray}
Using (\ref{pqprod}) and (\ref{eqjac}), we may also write
\begin{eqnarray}
\label{epqder}
E_{(p,q)}(z)&=& \sum_{n=0}^{\infty}
\left(\frac{q}{p}\right)^{n(n-1)/2}
\frac{z^{n}}{p^{-n(n+1)/2}(pq;pq)_{n}}
\cr
&=&
\sum_{n=0}^{\infty}
q^{n(n-1)/2} \frac{(zp)^{n}}{(pq;pq)_{n}}=E_{pq}(pz).
\end{eqnarray}
Then taking into account (\ref{invert}), (\ref{inv}) and (\ref{epqder}),
a $(p,q)$-analogue of (\ref{invert}) is given by
\begin{equation}
E_{pq}(-pz)e_{pq}(pz)= E_{(p,q)}(-z)e_{(p,q)}(p^{1/2}z)=1.
\end{equation}
Finally, consider
\begin{eqnarray}
\label{pqpens}
\mathfrak{e}_{(p,q)}^{(\mu,\nu)}(z)=
\sum_{n=0}^{\infty}\left(\frac{q^{\mu}}{p^{\nu}}\right)^{n^{2}}
\frac{z^{n}}{n!}.
\end{eqnarray}
Therefore, $\mathfrak{e}_{(p,q)}^{(\mu,\nu)}(z)$, which converges to $e^{z}$
as $(p,q)\to (1,1)$, provides a $(p,q)$-deformed exponential
analogue to the $q$-function used by Penson and Solomon \cite{pens2}
which coincides with $\mathfrak{e}_{(1,q)}^{(1,\nu)}(q^{-1/2}z)$.
The radius of convergence of (\ref{pqpens}) is given as
\begin{eqnarray}
\label{rad2}
R_2= \left\{\begin{array}{ll}
\infty, & \qquad\ {\rm if}\ q^{\mu}p^{-\nu} \leq 1; \\
0,  & \qquad\ {\rm if}\  q^{\mu}p^{-\nu} > 1.
\end{array}\right.
\end{eqnarray}

Finally, consider the Ramanujan integral \cite{ram,bal1},
valid for any integer $n\in\mathbb{N}$,
\begin{eqnarray}
\label{eq:App-Ramanujan}
\int_{0}^{\infty}dt\,t^n\,e_{q}(-t)
= - \frac{(q;q)_{n}}{q^{n(n+1)/2}} \log q.
\end{eqnarray}
Through the change of variables
\begin{eqnarray}
q \to pq ,\qquad t \to \lambda_0\,p^{-1/2}\,t,\qquad \lambda_0 >0,
\end{eqnarray}
and using once again (\ref{pqprod}), the following identity
is obtained, for any $n\in\mathbb{N}$,
\begin{eqnarray}
\int_{0}^{\infty} dt\,t^n\,e_{(p,q)}\left(-\lambda_0 p^{-1/2}t\right)
= \frac{[p,q;p,q]_{n}}{\lambda^{n+1}_0\,q^{n(n+1)/2}} \log\left(\frac{1}{pq}\right).
\label{eq:App-Ramanujan2}
\end{eqnarray}
This result is indeed a $(p,q)$-analogue of the Ramanujan integral
(\ref{eq:App-Ramanujan}).

\end{document}